# Finite-temperature properties of the extended Heisenberg model on a triangular lattice

P. Prelovšek[1,2] and J. Kokalj[3,1]

[1]*Jožef Stefan Institute, SI-1000 Ljubljana, Slovenia*
[2]*Faculty of Mathematics and Physics, University of Ljubljana, SI-1000 Ljubljana, Slovenia*
[3]*Faculty of Civil and Geodetic Engineering, University of Ljubljana, SI-1000 Ljubljana, Slovenia*

We present numerical results for the $J_1$-$J_2$ Heisenberg model on a triangular lattice at finite temperatures $T > 0$. In contrast to unfrustrated lattices we reach much lower $T \sim 0.15 J_1$. In static quantities the novel feature is a quite sharp low-$T$ maximum in the specific heat. Dynamical spin structure factor $S(\mathbf{q}, \omega)$ allows for the extraction of the effective spin-wave energies $\omega_\mathbf{q}(T)$ and their damping $\gamma_\mathbf{q}(T)$. While for $J_2 = 0$ our results are consistent with $T = 0$ spin ordering, $J_2/J_1 \sim 0.1$ induces additional frustration with a signature of spin liquid ground state. In the latter case, results for spin-lattice relaxation rate indicate in the low-$T$ accesible regime on $1/T_1 \propto T^\alpha$ with $\alpha \geqslant 1$, as observed in recent spin-liquid materials on a triangular lattice.

*Introduction.* In the last decade there are intensive theoretical efforts to understand spin systems, which would behave at low temperatures $T$ as spin liquids (SL) [1], i.e., revealing no long-range order (LRO) down to $T \to 0$. The theoretical search [2] has been stimulated by several novel materials, which seem to be realization of frustrated two-dimensional (2D) spin systems and experiments allow for the measurements of various static and dynamical properties [3]. As a very recent candidate with a triangular lattice (TL) is a layered material 1T-TaS$_2$, which has at $T < T_{CDW}$ a commensurate charge-density-wave (CDW) order, but is a Mott insulator [4] without any spin order at $T \to 0$ [5–8]. Nevertheless, the nuclear-relaxation measurements reveal spin fluctuations within the charge gap and spin-lattice relaxation $1/T_1 \propto T^\alpha$ with $\alpha \sim 2$ in a broad range $T_f < T < T_{CDW}$. Analogous relaxation is observed in organic compound $\kappa$-(ET)$_2$Cu$_2$(CN)$_3$ [3, 9, 10] with $\alpha \sim 1.5$ and EtMe$_3$Sb[Pd(dmit)$_2$]$_2$ with $\alpha \sim 2$ [11]. All these results indicate on relevant low-energy spin fluctuations and support the interpretation in terms of a SL, behaving as gapless at least in a certain $T$ range.

Antiferromagnetic (AFM) Heisenberg model (HM) on TL has been originally proposed [12] as the candidate for the resonant-valence-bond or SL ground state (g.s.). Numerical [13–15] and analytical [16] studies, however, reveal that g.s. has spiral LRO with spins pointing along $120^0$ tilted directions. Recently, the focus shifted to generalized HM where SL can become the stable g.s.. Recent numerical studies reveal that the $J_1 - J_2$ HM on TL with additional AFM next-neighbor coupling $J_2 > 0$ introduces further frustration which favours SL [17–22]. The same state might be stable also by introducing ring-exchange [23, 24] or similar terms obtained by downfolding the Hubbard model in the Mott insulating phase [25, 26]. Much less firm are conclusions on $T > 0$ properties. It is plausible that within a simple HM on TL frustration reduces LRO relative to unfrustrated square lattice [27]. Still, numerical calculations of static quantities within the HM on TL [28] did not pin down evident $T > 0$ features with a dominant effect of SL physics. In particular, the consequences for local spin correlations $S_L(\omega)$, as measured via NMR spin-lattice relaxation $1/T_1$, remain an open theoretical question in frustrated spin systems [3], but not fully clarified on the square lattice either [27, 29–31].

In this Letter we present $T > 0$ results for the $J_1 - J_2$ HM on TL obtained by the finite-$T$ Lanczos method (FTLM) [32–34]. One feature (not present on a square lattice) appears in the specific heat $C_V(T)$, revealing a broad shoulder at $T < J_1$ ending with a pronounced peak at $T_{coh} \sim 0.15 - 0.25 J_1$, least pronounced for the SL candidate at $J_2 \sim 0.1 J_1$. One can associate such a behavior with the onset of spin coherence (long- or short-range) only at $T \sim T_{coh} \ll J_1$. At $J_2 = 0$ we observe the softening of the 'spin-wave' (SW) energies $\omega_\mathbf{q}(T)$ as $T \to 0$ in the Brillouin zone (BZ) corner, consistent with the g.s. LRO [13–15] and the SW approximation [16]. However, small $J_2/J_1 \sim 0.1$ already induces near degeneracy at the BZ edges, favouring the interpretation in terms of SL. The most elusive conclusion so far has been the dynamical SW damping $\gamma_\mathbf{q}(T)$, which we find at lowest reachable $T$ decreasing with $T$. Consistently, we find in the SL regime at low but reachable $T$ above finite size $T_{fs} \sim 0.15 J_1$ the relaxation rate behaving as $1/T_1 \propto T^\alpha$ with $\alpha \geqslant 1$, qualitatively in agreement with recent results in candidate SL materials [3, 5].

*Model and numerical method.* We consider the isotropic $J_1$-$J_2$ extended HM on TL,

$$H = J_1 \sum_{\langle ij \rangle} \mathbf{S}_i \cdot \mathbf{S}_j + J_2 \sum_{\langle\langle ij \rangle\rangle} \mathbf{S}_i \cdot \mathbf{S}_j, \qquad (1)$$

for local spins $S = 1/2$ with AFM nearest-neighbor spin exchange $J_1$ and next-nearest neighbor exchange $J_2 \geqslant 0$. $J_2$ introduces additional frustration. Further on we set $J_1 = 1$ as an energy scale. We consider the model on 2D lattices with periodic boundary conditions with up to $N = 30$ sites according to minimal imperfection [35–37] and the possibility to contain particular wavevector

**Q** within the BZ. Within the triangular lattice of particular interest are BZ corner $\mathbf{Q}_K = (4\pi/3, 0)$ relevant for the $120^0$ spiral LRO, $\mathbf{Q}_M = (0, 2\pi/\sqrt{3})$ dominant in the stripe-like ordering at $J_2 \geqslant 0.15$ [21] or for square-lattice-like AFM order on anisotropic TL [24].

We calculate thermodynamic quantities and dynamical structure factor $S(\mathbf{q}, \omega)$ at $T > 0$, using the FTLM [32, 33], previously used in numerous studies of $T > 0$ properties in various models of correlated electrons [34]. The main limitation of the method are finite size effects appearing for $T < T_{fs}$. While for HM on a square lattice we reach $T_{fs} \sim 0.4$ [33, 34], the frustration due to TL and $J_2 > 0$ allows to reach considerably lower $T_{fs} \sim 0.15$. See more details in [37] where for example the specific heat $C_V(T)$ showing consistent behavior for a wide range of lattices $N = 20 - 30$ down to $T \sim 0.15$ is presented We note also that FTLM has been recently used also to evaluate $T > 0$ properties of HM and Hubbard model on an anisotropic TL [38, 39].

*Thermodynamic quantities.* The difference between the HM on a square lattice and on TL is well visible already in the entropy density $s(T)$ (in units of $k_B$) and in the related specific heat $C_V(T)$. Corresponding results for $J_2 = 0, 0.1, 0.2$, along with results for the square lattice, as obtained by FTLM on $N = 30$ lattice, are presented in Fig. 1. While in both models the entropy $s(T)$ saturates at $T > 1$, the behavior evidently splits at $T < 1$. On square lattice the onset of short-range order at $T < 1$ can be associated with a drop of $s(T)$, coinciding with the maximum in $C_V$ at $T = T_{coh} \sim 0.7$. On the other hand, large $s(T)$ persists on TL to lower $T$ and the inflection point shifts to $T_{coh} \sim 0.2$ for $J_2 = 0$ and even lower for $J_2 = 0.1, 0.2$. The behavior is reflected in $C_V(T)$ on TL, which reveals (in contrast to square lattice) a broad shoulder at $T < 1$, but as well a peak appearing at lower $T_{coh} = 0.15 - 0.25$. The latter is most pronounced and stable [37] at $J_2 = 0$, consistent with the onset of LRO in this case and in agreement with increased static correlations $S_\mathbf{q}$ and susceptibilities $\chi_\mathbf{q}^0$ for $T < T_{coh}$ [37]. It has been apparently missed in some previous studies [28], but observed within Hubbard model on the anisotropic TL [39]). The peak in $C_V(T)$ is weakest for $J_2 \sim 0.1$, another signature for low-$T$ SL, while it reappears for $J_2 = 0.2$ with the presumable onset of the stripe order. In contrast to $C_V(T)$ the uniform susceptibility $\chi_0(T)$ is more featureless [37].

*Dynamical spin structure factor.* Next we turn to the dynamical spin structure factor $S(\mathbf{q}, \omega)$, evaluated within FTLM [33, 34] with similar restrictions as the thermodynamic quantities above. Due to periodic boundary conditions we are restricted to discrete $\mathbf{q}$ within the BZ (see the inset in Fig. 2 for the $N = 30$ lattice). It is helpful to analyse $S(\mathbf{q}, \omega)$ in terms of the dynamical susceptibilities $\mathrm{Im}\chi_\mathbf{q}(\omega) = \pi(1 - e^{-\beta\omega})S(\mathbf{q}, \omega)$ whereby (complex) $\chi_\mathbf{q}(\omega)$ can be represented in the form of the memory function

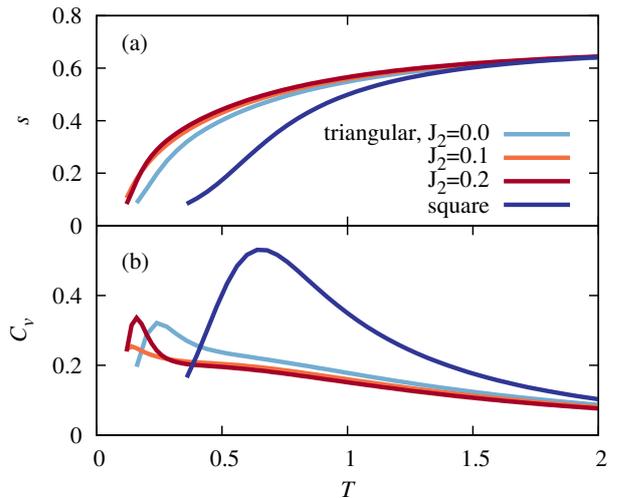

Figure 1. ( a) The entropy density $s(T)$, and b) specific heat $C_V(T)$ of the $J_1$-$J_2$ Heisenberg model on the triangular lattice for different $J_2$, and compared to the reference square-lattice result, as obtained by FTLM on lattices with $N = 30$ sites.

(MF) $M_\mathbf{q}(\omega)$ [40–42] (for relations see [37]),

$$\chi_\mathbf{q}(\omega) = \frac{-\eta_\mathbf{q}}{\omega^2 - \omega_\mathbf{q}^2 + \omega M_\mathbf{q}(\omega)}, \qquad (2)$$

with the effective SW energy $\omega_\mathbf{q}^2 = \eta_\mathbf{q}/\chi_\mathbf{q}^0$, $\chi_\mathbf{q}^0 = \chi_\mathbf{q}(\omega = 0)$ and $\eta_\mathbf{q}$ the corresponding stiffness. Both $\chi_\mathbf{q}^0$ and $\eta_\mathbf{q}$ can be directly evaluated from numerically calculated $S(\mathbf{q}, \omega)$, while the latter allows also to extract $M_\mathbf{q}(\omega)$ [37]. We note that $\gamma_\mathbf{q}(\omega) = \mathrm{Im}M_\mathbf{q}(\omega)$ has a direct interpretation in terms of the damping of the SW mode at $\omega_\mathbf{q}$. The analysis in terms of Eq. (2) is general, but it becomes particularly meaningful provided that $\gamma_\mathbf{q}(\omega)$ is weakly $\omega$ and $\mathbf{q}$ dependent, especially for $\mathbf{q}$ along the BZ boundary, where $\omega_\mathbf{q}$ are smallest (and $\chi_\mathbf{q}^0$ largest). We note that $\eta_\mathbf{q}$ (expressed in terms of local static correlations [37]) is (away from $q = 0$) generally modestly $\mathbf{q}$ and $T$ dependent.

Results for $\chi_\mathbf{q}^0$, presented in Fig. 2a, reveal crucial dependence on $J_2$. We compare $\chi_\mathbf{q}^0(T)$ on TL for three parameters $J_2 = 0, 0.1, 0.2$ and for three characteristic $\mathbf{q}$: BZ corner $\mathbf{Q}_K$ and edge $\mathbf{Q}_M$, and for reference also the inside-BZ point $\mathbf{Q}_X$ (where SW approximation at $J_2 = 0$ yields the maximum $\omega_\mathbf{q} \sim 1.5$ [16]), which are accesible on $N = 30$ lattice (see inset in Fig. 2). Several conclusions can be drawn from Fig. 2a: a) $\chi_\mathbf{q}^0(T)$ have strong $T$ dependence and much larger values for $\mathbf{q}$ along the BZ boundaries relative to inside of BZ (as well to $\chi_0(T)$ [37]). b) The variation with $J_2$ shows the qualitative change of spin excitations. For $J_2 = 0$ the value at $\mathbf{Q}_K$ is clearly dominant over $\mathbf{Q}_M$, consistent with divergence of $\chi_K^0(T \to 0)$ and the spiral g.s. LRO. This is not the case for $J_2 = 0.1$ where $\chi_M^0(T) \sim \chi_K^0(T)$ with an increase for lower $T \sim T_{fs}$, whereby we cannot give a firm conclu-sion on the behavior of $\chi_\mathbf{q}^0(T \to 0)$. Finally, for $J_2 = 0.2$ the situation is inverted relative to $J_2 = 0$, with dominant and apparently diverging $\chi_M^0(T \to 0)$, indicating a stripe g.s. LRO. c) Presented results for $\chi_\mathbf{q}(T \to 0)$ are qualitatively consistent with previous numerical studies of the g.s. of the $J_1$-$J_2$ HM on TL, dealing with static (equal-time) correlations $S_\mathbf{q} = (1/\pi) \int d\omega S(\mathbf{q},\omega)$ [17, 19, 21, 22] which also show a qualitative change along the $J_2$ phase diagram (for our $S_\mathbf{q}(T)$ results see [37]). In contrast to $S_\mathbf{q}(T)$ the susceptibility $\chi_\mathbf{q}^0(T)$ incorporates also information on dynamical softening, and enhances differences between low-$T$ regimes.

Corresponding $\omega_\mathbf{q}(T)$ are presented in Fig. 2b and mainly reflect the behavior of $\chi_\mathbf{q}^0(T)$. Results are qualitatively consistent with the softening of SW mode $\omega_K(T \to 0) = 0$ at $J_2 = 0$, and similarly $\omega_M(T \to 0) = 0$ for $J_2 = 0.2$, respectively. On contrary, consistent with the SL scenario $\omega_\mathbf{q}(T \to 0)$ only partially softens in the intermediate case $J_2 = 0.1$ with near degeneracy of $\omega_K \sim \omega_M$ down to lowest $T$. Results for $\omega_\mathbf{q}(T)$ could be qualitatively interpreted in terms SW-approximation result $\tilde\omega_\mathbf{q}$ [16] generalized via an effective gap $\kappa(T)$ (at least along the BZ boundary) into $\omega_\mathbf{q}^2 \sim \tilde\omega_\mathbf{q}^2 + \kappa^2$, which can as well be associated with the finite AFM correlation length $\xi \sim v_0/\kappa$, where $v_0 \sim 1$ is the AFM magnon velocity [16] (at $J_2 = 0$). In such a scenario, results confirm that in HM on TL the longer correlation length, i.e., $\kappa(T) < 1$ emerges only at $T < 0.5$, below which $\omega_\mathbf{q}$ start to differentiate. I.e., for $J_2 = 0$ $\omega_K(T)$ is decreasing with $T \to 0$, while $\omega_X$ increases towards the SW-approximation result $\omega_X \sim 1.6$ [16]. The role of $K$ and $M$ are reversed for $J_2 = 0.2$, while most interesting $J_2 = 0.1$ reveals besides the degeneracy $\omega_K \sim \omega_M$ and $\kappa(T) > 0.7$ also weak $T$ dependence, at least for reachable $T \sim T_{fs}$.

Although one might suspect that damping $\gamma_\mathbf{q}(\omega)$ is more sensitive quantity (which is anyhow the case due to numerical and finite-size limitations), results confirm that it is (some characteristic cases shown in [37]) quite $\omega$ independent, at least within the relevant range $\omega \lesssim \omega_\mathbf{q}$. This reduces Eq. (2) to a form of simple damped oscillator [37]. In Fig. 3 we show results for the d.c. limit $\gamma_\mathbf{q}^0 = \gamma_\mathbf{q}(\omega \to 0)$ which reveal that it is moreover weakly $\mathbf{q}$ dependent. At high $T > 0.5$ we get $\gamma_\mathbf{q}^0 \sim 2$, in comparison to $\omega_\mathbf{q} \gtrsim 1$ in Fig. 2b confirming well damped excitations within the whole BZ. For $T < 0.5$ $\gamma_\mathbf{q}^0$ decreases more strongly than $\omega_q$ and leads to more underdamped excitations in particular for $\mathbf{Q}_K$ at $J_2 = 0$ and $\mathbf{Q}_M$ at $J_2 = 0.2$.

*NMR spin-lattice relaxation.* For the interpretation of NMR spin-lattice relaxation time $T_1$ the important relation is with low $\omega$ dynamical local spin correlations $S_L(\omega) = (1/N)\sum_\mathbf{q} S(\mathbf{q},\omega)$. Assuming $\mathbf{q}$-independent form factor it follows from Eq. (2),

$$\frac{1}{T_1} \propto S_L(\omega \to 0) = \frac{T}{\pi N}\sum_\mathbf{q} \frac{\eta_\mathbf{q}\gamma_\mathbf{q}^0}{\omega_\mathbf{q}^4}. \quad (3)$$

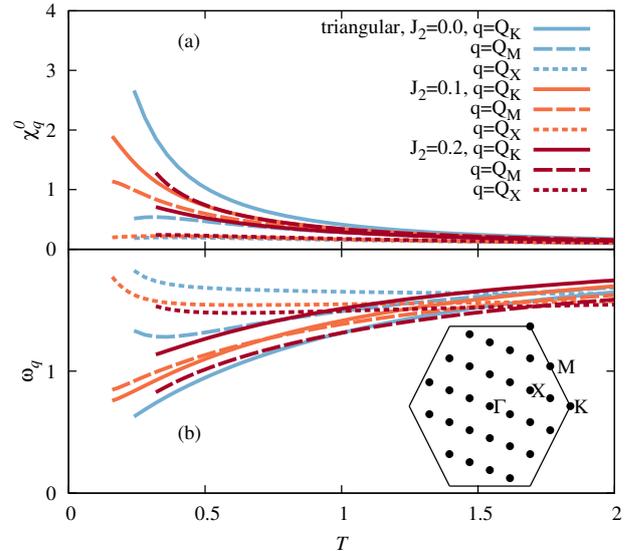

Figure 2. a) Static susceptibilities $\chi_\mathbf{q}^0(T)$, and b) corresponding $\omega_\mathbf{q}(T)$ on TL for different $J_2 = 0, 0.1, 0.2$, presented for different $\mathbf{q}$: corner $K$, edge $M$ and inside $X$ BZ points, respectively. Inset shows the BZ and available $\mathbf{q}$ on $N = 30$ lattice.

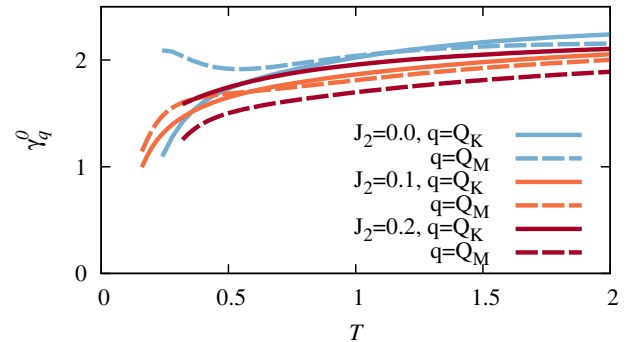

Figure 3. D.c. damping $\gamma_\mathbf{q}^0(T)$ for different $J_2 = 0, 0.1, 0.2$ at two BZ-edge wavevectors for $N = 30$ lattice.

Results of a straightforward FTLM calculation of $1/T_1$ obtained on $N = 30$ system (omitting the $q = 0$ contribution) are presented in Fig. 4 for the TL with different $J_2$, along with the corresponding result for the square lattice. Using Eq. (3), results can be interpreted as follows. For $T > 1$ $S_L(\omega \sim 0)$ becomes $T$-independent, so we get $1/T_1 \sim$ const. [43], with the value quite insensitive to lattice as well as on $J_2$. Such behavior extends for TL even further to $T > 0.5$. At lower $T$ the $T$ variation enters through Eq. (3) via $\omega_\mathbf{q}(T)$ and $\gamma_\mathbf{q}(T)$, where the dominant contribution to $S_L(\omega)$ emerges from soft modes $\omega_\mathbf{q} < 1$ inside the BZ. One can simplify the discussion with the observation that the damping is quite $\mathbf{q}$ independent, i.e., $\gamma_\mathbf{q}^0 \sim \gamma^0$, while $\eta_\mathbf{q} = \eta^0$ is, moreover, $T$ inde-pendent. From Eq. (3) we then arrive at the estimate for the $T$-dependence as $1/T_1 \propto T\gamma^0(T)S(T)/\kappa^4(T)$, where $S(T)$ is the BZ surface satisfying $\omega_\mathbf{q} \sim \kappa$. Several different scenarios can be now anticipated from the preceding relation for low $T$:

a) In the case of g.s. LRO in 2D the divergent AFM correlation length is expected as $\kappa = 1/\xi \propto e^{-A/T}$ [44]. Since in 2D $S \propto \kappa^2$ and diverging $\xi(T \to 0)$ dominates over vanishing $\gamma^0 \propto T^\zeta$, this leads finally to diverging $1/T_1(T \to 0) \to \infty$ [29–31]. Due to finite size (restricted number of $\mathbf{q}$) this behavior is quite hard to establish numerically. Still, we find a reproducible upturn in agreement with such a scenario in Fig. 4, both for the square lattice with an upturn at $T \sim 0.7$, as well in the TR lattice with $J_2 = 0$ with the minimum of $1/T_1$ at $T \sim 0.5$. From our results in Fig. 4 it is not clear whether an analogous scenario should finally (at $T \to 0$) also apply to stripe order at $J_2 = 0.2$.

b) In a SL regime spin LRO does not emerge at $T \to 0$. This is reflected either in the saturation $\kappa(T \to 0) \sim \kappa^0$ or very slowly varying and possibly vanishing $\kappa(T \to 0) \to 0$. In the range of our method it is hard to distinguish both options, nevertheless one of them appears to be the case at intermediate $J_2 \sim 0.1$. This still offers several scenarios for the SL. Finite limiting $\gamma^0(T \to 0)$, with $\kappa_0 > 0$ would indicate on fermionic-type excitations and a gapless SL with spinon Fermi surface [7, 8], leading to $1/T_1 \propto T$, not excluded by results on Figs. 2b and 3. The vanishing $\gamma^0(0)$, which appears more plausible from Fig. 3, at finite $\kappa_0 > 0$ would imply a gapped SL. In this case, observed power-law-like $1/T_1 \propto T^\alpha$ with $\alpha \geqslant 1$ at $T < T_{coh}$ would crossover into an low-$T$ exponential dependence. On the other hand, slowly vanishing $\kappa(T \to 0) \to 0$ would allow also for more involved scenarios, as the gapless Dirac SL, which would lead to $\alpha = 2$ [7], not inconsistent with Fig. 4.

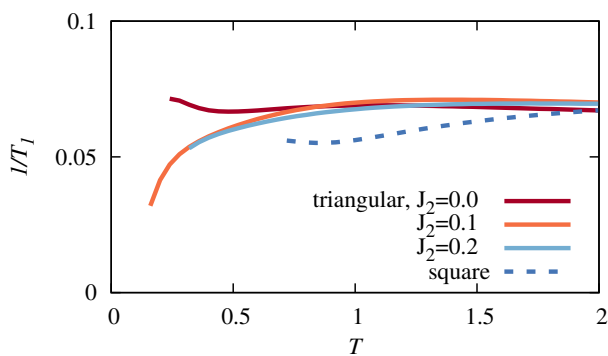

Figure 4. Spin-lattice relaxation rate $1/T_1$ vs. $T$ for the HM on TL with different $J_2 = 0 - 0.2$. For comparison the result for the square lattice is also shown.

*Conclusions.* Presented results of $T > 0$ static and dynamical properties of 2D $J_1$-$J_2$ HM on TL are com-plementary to several established and recent studies of the g.s. of the same model. The comparison with better understood HM on square lattice is quite informative. One evident difference is that relative to latter case, the frustration allows for reliable results to much lower, i.e., $T_{fs} \sim 0.15 J_1$ for the most frustrated case $J_2 \sim 0.1$. In the specific heat $C_V(T)$ this enables to establish (in contrast to square lattice) two characteristic scales. Besides the broad shoulder at $T < J_1$, we find in TL quite sharp peak at $T \sim (0.15 - 0.2)J_1$ indicating the onset of spin coherence, pronounced in particular for $J_2 = 0$ case.

The focus is still on $T > 0$ dynamical properties, not much accessible so far, in particular those relevant for recent experiments on SL. For the $J_2 = 0$ we find softening of the BZ corner mode $\omega_K(T \to 0) \to 0$ [13–15] as well as expected increasing NMR relaxation rate $1/T_1$ by lowering $T$, as expected for systems with AFM LRO [29, 30]. Most interesting is clearly the intermediate regime $J_2/J_1 \sim 0.1$ as the candidate for SL. Here, down to lowest $T$ a near degeneracy along the BZ edge remains, i.e. $\omega_K \sim \omega_M$ showing tendency of finite limiting $T \to 0$ value in for low $T$ regime, e.g., for $T_{fs} < T < T_{coh}$. Together with decreasing damping $\gamma_0(T \to 0)$, this leads to observed vanishing $1/T_1 \propto T^\alpha$ with $\alpha \geqslant 1$ in the lowest reachable $T$ regime. Such a behavior is an experimental hallmark of a SL, established with $\alpha \sim 2$ in several recent experiments [3, 5]. Still, from our study it is hard to distinguish whether this is intermediate $T$ dependence within a possible gapped SL, or an indication of a gapless SL, e.g., with spinon Fermi-surface with $\alpha = 1$, or a Dirac SL with $\alpha = 2$.

This work is supported by the program P1-0044 of the Slovenian Research Agency.

# Supplemental Material for: Finite-temperature properties of the extended Heisenberg model on a triangular lattice

P. Prelovšek[1,2] and J. Kokalj[3,1]

[1]*Jožef Stefan Institute, SI-1000 Ljubljana, Slovenia*
[2]*Faculty of Mathematics and Physics, University of Ljubljana, SI-1000 Ljubljana, Slovenia*
[3]*Faculty of Civil and Geodetic Engineering, University of Ljubljana, SI-1000 Ljubljana, Slovenia*

In the Supplemental Material we provide more details concerning the numerical method and tests, as well some additional results on thermodynamical quantities, the basic facts on the memory-function analysis and the presentation of dynamical quantitites.

## I. NUMERICAL METHOD AND THERMODYNAMICAL QUANTITIES

In the evaluation of thermodynamical (and further on dynamical) quantities we employ the FTLM method [1–3] on various triangular (and for comparison also square) lattices with periodic boundary conditions. The lattices with $N = 20 - 30$ sites are chosen so that they have a small imperfection [4, 5], but as well that they contain (if possible) characteristic points in the Brillouin zone. In particular this is the case for $\mathbf{Q}_K = (4\pi/3, 0)$ (the ordering wavevector of the spiral 120° order in triangular lattice) and $\mathbf{Q}_M = (0, 2\pi/\sqrt{3})$, ordering for the stripe-like order for $J_2 > 0.15$ and for the related standard AFM order on square lattice obtained as an extreme limit of anisotropic triangular lattice. Among the presented results, the largest lattice with $N = 30$ contains both points (see inset in Fig. 2 in the main text), while smaller lattices contain only some of them, e.g., $N = 26$ contains $\mathbf{Q}_M$ but not $\mathbf{Q}_K$. Similar is the choice in presented results for the HM on the square lattice with $N = 30$ sites which contains the AFM ordering wavevector $\mathbf{Q}_M = (\pi, \pi)$.

The limitation of the present method is given by the size of the many-body Hilbert space with $N_{st} < 10^7$ basis states which can be handled efficiently within the FTLM, restricting the lattice sizes to $N \leq 30$. The main criterion for the macroscopic relevance of FTLM results is the thermodynamical sum $Z(T) = \text{Tr}\exp[-(H - E_0)/T]$, which determines the $Z > Z^* = Z(T_{fs}) \gg 1$. In addition, the FTLM approach also involves a random sampling over initial wavefunctions, which we typically choose $N_s = 10$, which partly influences the statistical error [2, 3] of results. We use the actual criterion that $Z^* \sim 10$ which implies (for $N = 30$) also roughly the threshold at entropy density $s(T_{fs}) \sim 0.1$, independent of the model. It is then evident that from Fig. 1 in the main text that $T_{fs}$ depends essentially on the model. In particular, it follows that $T_{fs} \sim 0.4J$ for the square lattice HM, while $T_{fs} \sim 0.2J$ for the triangular lattice and $T_{fs} \sim 0.15J$ for frustrated lattice with $J_2 = 0.1$. While results at higher $T > T_{fs}$ do depend on the choice of the lattice, the actual behavior of various quantities, in particular of $\mathbf{q}$-dependent or $\mathbf{q}$-integrated dynamical quantities as $1/T_1$, is more sensitive on the presence of relevant symmetry point in the Brillouin zone. On the other hand, thermodynamic quantities as $s(T)$, $C_V(T)$ and $\chi_0(T)$ are more robust, so results remain quite lattice-independent even at lower $T < T_{fs}$.

In Fig. 1 we present the results for the specific heat $C_V(T)$ on a triangular lattice with n.n. exchange only, i.e. with $J_2 = 0.$, as obtained on different lattices with $N = 20 - 30$. It is evident that qualitative features of $C_V(T)$ are quite consistent and size independent down to $T \sim 0.15J$. All results share two essential features: a broad shoulder appearing below $T \sim J$ and a narrow coherence peak at $T \sim 0.2J$. Here, one should acknowledge that among presented thermodynamic quantities, $C_V$ is the most sensitive on finite size effects. Hence, final results for $C_V(T)$ as well as for the entropy $s(T)$ for other model parameters as presented in Fig. 1 in the main text can be regarded as macroscopic ones down to $T \sim T_{fs}$.

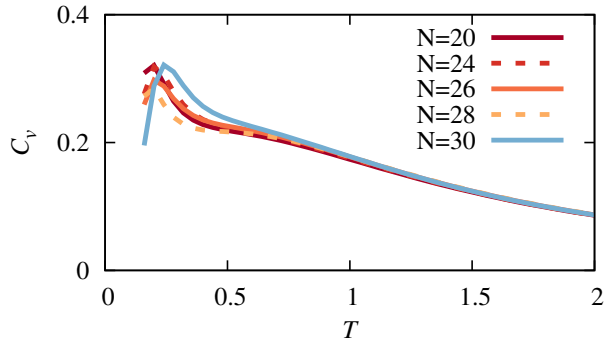

FIG. 1. Results for the specific heat $C_V(T)$ of the Heisenberg model on triangular lattice with $J_2 = 0$, as obtained by FTLM on systems with different number of sites $N = 20 - 30$.

In Fig. 2 we present also results for the uniform susceptibility $\chi_0(T)$. For the triangular lattice $\chi_0(T)$ are only weakly dependent on $J_2$, at least for the accesible range $T > 0.15J$, since they all reveal maximum at $T^* \sim 0.4J$, in agreement (both in location and the value) with the high-$T$ expansion result [6]. For comparison we show also the result for the square lattice with the maximum at $T^* \sim J$. While in the latter case the low-$T$ limit is understood (due to long-range order at $T = 0$ one has

$\chi_0^0 = \chi_0(T \to 0) > 0$) [7], this is not the case for the frustrated triangular lattice with finite $J_2$. In the case of $J_2 = 0$ (pure triangular lattice) one would expect in analogy with the square lattice $\chi_0^0 > 0$, while in the SL case, e.g., at $J_2 = 0.1$, the spin excitations can be either gapped with $\chi_0^0 = 0$ or gapless with $\chi_0^0 > 0$. Still, the restriction to $T > T_{fs}$ does not allow to differentiate between possible limits (finite or zero) of $\chi_0(T \to 0)$.

For completeness we show in Fig. 3 also the static spin structure factor $S_\mathbf{q}$ for chosen $q$'s in the Brillouin zone.

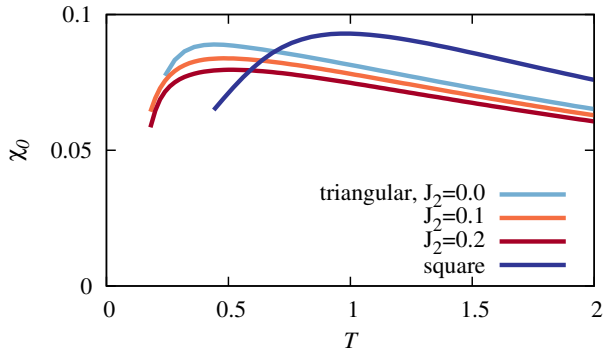

FIG. 2. Uniform spin susceptibility $\chi_0(T)$ within the Heisenberg model on the triangular lattice with $J_2 = 0 - 0.2$ and on the square lattice. Obtained by FTLM on $N = 30$ lattice.

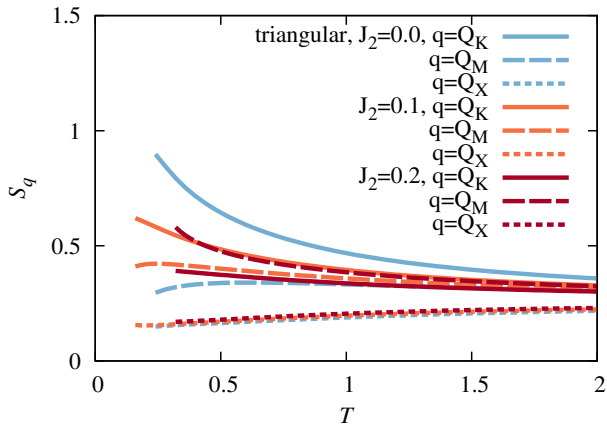

FIG. 3. Static spin structure factor $S_\mathbf{q}(T)$ for several wavevectors $\mathbf{q}$ calculated within the Heisenberg model on the triangular lattice with $J_2 = 0 - 0.2$. Obtained by FTLM on $N = 30$ lattice.

## II. MEMORY-FUNCTION ANALYSIS

The dynamical spin susceptibility $\chi_\mathbf{q}(\omega)$ and the dynamical structure factor $S(\mathbf{q},\omega)$ are defined in a standard way,

$$\chi_\mathbf{q}(\omega) = i\int_0^\infty dt e^{i\omega^+ t}\langle[S_\mathbf{q}^z(t), S_{-\mathbf{q}}^z(0)]\rangle,$$
$$\mathrm{Im}\chi_\mathbf{q}(\omega) = \pi(1-e^{-\beta\omega})S(\mathbf{q},\omega), \quad (1)$$

The memory function (MF) formalism [8–10] is as usual introduced via the corresponding relaxation function $\Phi_\mathbf{q}(\omega)$,

$$\Phi_\mathbf{q}(\omega) = \frac{\chi_\mathbf{q}(\omega) - \chi_\mathbf{q}^0}{\omega}, \quad (2)$$

where $\chi_\mathbf{q}^0 = \chi_\mathbf{q}^0(\omega = 0)$ is static spin susceptibility. Then, one can generally represent the complex $\Phi_\mathbf{q}(\omega)$ in terms of the effective SW energy $\omega_\mathbf{q}$ and the (complex) SW damping function $M_\mathbf{q}(\omega)$,

$$\Phi_\mathbf{q}(\omega) = \frac{-\chi_\mathbf{q}^0}{\omega - \frac{\omega_\mathbf{q}^2}{\omega + M_\mathbf{q}(\omega)}}, \qquad \omega_\mathbf{q}^2 = \frac{\eta_\mathbf{q}}{\chi_\mathbf{q}^0}, \quad (3)$$

leading with relation (2) to the Eq. (2) in the main text. In the formalism $\omega_\mathbf{q}$ is given as the (second) frequency moment of the relaxation function $\mathrm{Im}\Phi_\mathbf{q}(\omega)$ with $\eta_\mathbf{q}$ being the corresponding 'stiffness',

$$\eta_\mathbf{q} = \frac{1}{\pi}\int_{-\infty}^\infty \omega\,\mathrm{Im}\chi_\mathbf{q}(\omega)d\omega = \langle[[S_\mathbf{q}^z, H], S_{-\mathbf{q}}^z]\rangle \quad (4)$$

It is evident from the above relation to the static (rather local) quantity, that $\eta_\mathbf{q}$ is expected to be smoothly $\mathbf{q}$ dependent as well weakly $T$ dependent. From known $S(\mathbf{q},\omega)$ we can now directly evaluate static quantities $\chi_\mathbf{q}^0$ and $\omega_\mathbf{q}$ and consequently from the complex $\chi_\mathbf{q}(\omega)$ (obtained via the Hilbert transform) we extract the MF function $M_\mathbf{q}(\omega)$,

$$M_\mathbf{q}(\omega) = \frac{\omega_\mathbf{q}^2}{\omega + \chi_\mathbf{q}^0/\Phi_\mathbf{q}(\omega)} - \omega, \quad (5)$$

in particular the SW damping function $\gamma_\mathbf{q}(\omega) = \mathrm{Im}M_\mathbf{q}(\omega)$.

## III. DYNAMICAL QUANTITIES

The analysis of $S(\mathbf{q},\omega)$ in terms of the MF (3) is particularly meaningful, when the MF is considerably smoother, e.g., both in $\omega$ and in $\mathbf{q}$, than the corresponding original $S(\mathbf{q},\omega)$ or $\Phi_\mathbf{q}(\omega)$.

In Fig. 4 we show imaginary part of the relaxation function $\mathrm{Im}\Phi_\mathbf{q}(\omega)$, and its $\omega$ and $\mathbf{q}$ dependence can be compared to the $\omega$ and $\mathbf{q}$ of the memory function. In Fig. 5 we show imaginary part of MF $\gamma_\mathbf{q}(\omega)$, and demonstrate its weak $\omega$ and $\mathbf{q}$ dependence for a range of $T$.

For completeness we show in Fig. 6 imaginary part of local relaxation function

$$\frac{\mathrm{Im}\Phi_L(\omega)}{\pi} = \frac{1}{\pi N}\sum_\mathbf{q}\mathrm{Im}\chi_\mathbf{q}(\omega)/\omega, \quad (6)$$

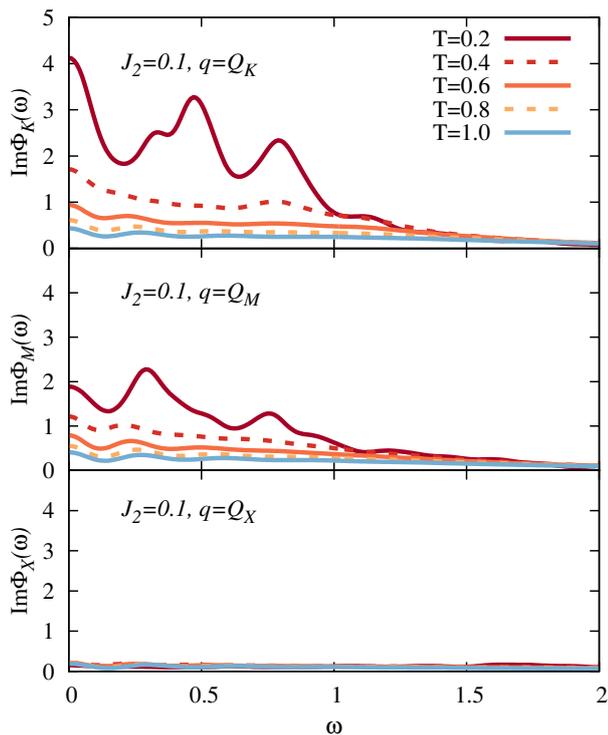

FIG. 4. Frequency dependence of the imaginary part of the relaxation function $\mathrm{Im}\Phi_\mathbf{q}(\omega)$ for three characteristic $\mathbf{q} = \mathbf{Q}_K$, $\mathbf{Q}_M$ and $\mathbf{Q}_X$. Calculated with FTLM on $N = 30$ lattice for $J_2 = 0.1$ and broadened with $\eta = 0.05$.

which is directly related to the NMR relaxation rate $1/T_1 = T\mathrm{Im}\Phi_L(\omega)/\pi$ and shows smooth $\omega$ dependence and a decrease with $T$.

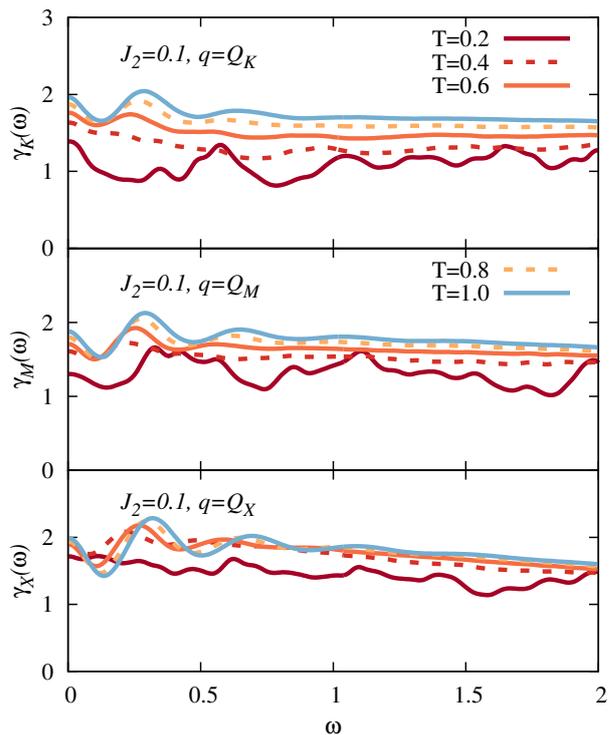

FIG. 5. Frequency dependence of the imaginary part of memory function, $\gamma_\mathbf{q}(\omega) = \mathrm{Im}M_\mathbf{q}(\omega)$, for three characteristic $\mathbf{q} = \mathbf{Q}_K$, $\mathbf{Q}_M$ and $\mathbf{Q}_X$. It is evident that $\gamma_\mathbf{q}(\omega)$ shows surprisingly small dependence on $\omega$ and $\mathbf{q}$, while it shows a decrease with $T$ at lowest $T$ as expected. Its value at $\omega = 0$ determines the NMR relaxation rate $1/T_1$. Calculated with FTLM on $N = 30$ lattice for $J_2 = 0.1$ and broadened with $\eta = 0.05$.

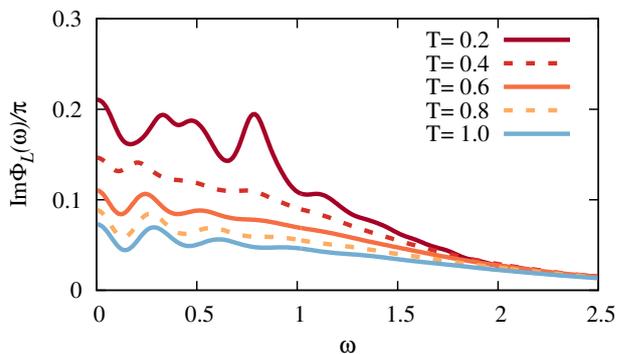

FIG. 6. Frequency dependence of the local $\mathrm{Im}\Phi_L(\omega)/\pi$ for several $T$. Calculated with FTLM on $N = 30$ lattice for $J_2 = 0.1$ and broadened with $\eta = 0.05$.